\documentclass[sigconf]{acmart}

\AtBeginDocument{%
  }

\newcommand{\add}[1]{#1}
\newcommand{\delete}[1]{}  

\renewcommand\footnotetextcopyrightpermission[1]{} 

\acmConference[ICSE '26]{}{Apr 14 -- 21,
  2026}{Rio De Janeiro, Brazil}

\usepackage{hyperref}
\usepackage{hyperxmp}
\usepackage[]{collab}
\usepackage{ulem}
\usepackage{enumitem}
\usepackage{balance}
\usepackage{xcolor}
\usepackage{setspace}
\usepackage{subfigure}
\usepackage{multirow}
\usepackage{multicol}
\usepackage{tabularx}
\usepackage{siunitx}
\usepackage{float}
\usepackage{color}
\usepackage{diagbox}
\usepackage{caption}
\usepackage{makecell}
\usepackage{colortbl}
\usepackage{tcolorbox}
\usepackage{mdframed}
\usepackage{listings}
\usepackage{booktabs}
\usepackage[utf8]{inputenc}
\usepackage{calligra}
\usepackage[normalem]{ulem}
\usepackage{indentfirst}
\usepackage{url}
\usepackage{soul}
\usepackage{nicematrix}
\usepackage{fontawesome5}
\usepackage{textcomp}
\usepackage{stfloats}
\usepackage{verbatim}
\usepackage{graphicx}
\usepackage{amsmath,amsfonts}
\usepackage[linesnumbered,ruled,vlined]{algorithm2e}
\usepackage{array}
\usepackage[utf8]{inputenc}

\definecolor{ballblue}{rgb}{0.13, 0.67, 0.8}
\definecolor{grey}{rgb}{0.9, 0.9, 0.9}
\definecolor{googlered}{rgb}{0.914, 0.262, 0.207}
\definecolor{dandelion}{rgb}{0.95, 0.65, 0.0}

\collabAuthor{jj}{blue}{Junjie}
\collabAuthor{gb}{magenta}{Guangba}
\collabAuthor{zh}{purple}{Zhihan Jiang}
\collabAuthor{jz}{magenta}{Jiazhen Gu}
\collabAuthor{zb}{teal}{Zhuangbin}
\collabAuthor{jc}{googlered}{JC}
\collabAuthor{yj}{orange}{Yujie Huang}

\newcommand\nm{PreServe\xspace}

\newcommand{\ie}{{\em i.e.},\xspace}
\newcommand{\eg}{{\em e.g.},\xspace}

\newcommand{\boxmargin}{1mm}

\newtcolorbox{myboxa}[2][]{
    colback=gray!10!white,
    colframe=black, enhanced,
    attach boxed title to top left={yshift=-2mm,xshift=5mm},
    title=#2,#1
}
\newtcolorbox{myboxb}[2][]{
    boxsep=3pt,
    left = \boxmargin, right = \boxmargin, top = \boxmargin, bottom = \boxmargin,
    title={#2},#1
}
\newtcolorbox{myboxc}{
    colback=gray!15!white,
    arc = 0pt, outer arc = 0pt,
    boxsep=0pt, left = 3pt, right = 0pt, top = 0pt, bottom = 0pt, 
    leftrule=3pt, bottomrule=0pt,toprule=0pt, rightrule=0pt,
    left = \boxmargin, right = \boxmargin, top = \boxmargin, bottom = \boxmargin
}
\newtcolorbox{myboxd}{
    colback=gray!10,
    colframe=black,
    width=\columnwidth,
    arc=1mm, auto outer arc,
    boxrule=0.5pt,
}

\definecolor{myyellow}{HTML}{FFF2CC}
\newcounter{finding}

\definecolor{myyellow}{HTML}{FFF2CC}
\newcounter{insight}

\newcounter{challenge}
\newcommand{\challenge}[1]{\refstepcounter{challenge}
 	\vspace{1mm}
	\begin{mdframed}[linecolor=gray!25,roundcorner=12pt,backgroundcolor=myyellow!20,linewidth=3pt,innerleftmargin=2pt, leftmargin=0cm,rightmargin=0cm,topline=false,bottomline=false,rightline=false,leftline=false]
		\textbf{Management Challenge \arabic{challenge}:} #1
	\end{mdframed}
}

\definecolor{mygreen}{HTML}{AFCFA5}
\newcounter{opportunity}
\newcommand{\opportunity}[1]{\refstepcounter{opportunity}
 	\vspace{1mm}
	\begin{mdframed}[linecolor=gray!25,roundcorner=12pt,backgroundcolor=mygreen!20,linewidth=3pt,innerleftmargin=2pt, leftmargin=0cm,rightmargin=0cm,topline=false,bottomline=false,rightline=false,leftline=false]
		\textbf{Management Opportunity \arabic{opportunity}:} #1
	\end{mdframed}
}

\begin{document}
\title{\nm: Intelligent Management for LMaaS Systems via Hierarchical Prediction}

\author{Zhihan Jiang}
\affiliation{%
  \institution{The Chinese University of Hong Kong}
  \city{Hong Kong SAR}
  \country{China}}

\author{Yujie Huang}
\affiliation{%
  \institution{The Chinese University of Hong Kong}
  \city{Hong Kong SAR}
  \country{China}}

\author{Guangba Yu}
\affiliation{%
  \institution{The Chinese University of Hong Kong}
  \city{Hong Kong SAR}
  \country{China}}

\author{Junjie Huang}
\affiliation{%
  \institution{The Chinese University of Hong Kong}
  \city{Hong Kong SAR}
  \country{China}}

\author{Jiazhen Gu}
\affiliation{%
  \institution{The Chinese University of Hong Kong}
  \city{Hong Kong SAR}
  \country{China}}

\author{Michael R. Lyu}
\affiliation{%
  \institution{The Chinese University of Hong Kong}
  \city{Hong Kong SAR}
  \country{China}}

\begin{abstract}


Large Language Models (LLMs) have revolutionized numerous domains, driving the rise of Language-Model-as-a-Service (LMaaS) platforms that process millions of queries daily.
These platforms must minimize latency and meet Service Level Objectives (SLOs) while optimizing resource usage. 
However, conventional cloud service management techniques, designed for traditional workloads, are suboptimal for LMaaS due to its dynamic service workloads and variable request loads.
To address this, we propose \nm, a tailored LMaaS management framework centered on hierarchical prediction.
\nm incorporates a \emph{service workload predictor} to estimate periodic token density at a coarse granularity and a novel \emph{request load predictor} to assess the resource demand of individual LLM requests, enabling the construction of a \emph{load anticipator} for each LLM instance.
By integrating both long-term and short-term predictions, \nm adjusts resource allocation in advance, mitigating the risks of instance under- or over-provisioning.
Besides, \nm optimizes request routing by considering both current and anticipated future instance loads, ensuring balanced load distribution across instances.
Evaluations on real-world production datasets show that \nm outperforms state-of-the-art methods, reducing tail latency by 41.3\%, cutting resource consumption by 49.38\%, while incurring only 0.23\% additional overhead.
\end{abstract}

\maketitle

\section{introduction}

Large language models (LLMs) have emerged as transformative technologies in various domains, such as natural language processing~\cite{he2023large,yang2024harnessing} and software engineering~\cite{gao2023makes,lyu2024automatic}.
Their applications in areas such as conversational systems~\cite{ChatGPT,DeepSeekChat}, search engines~\cite{xiong2024searchengineservicesmeet,chen2024mindsearchmimickinghumanminds}, and code generation~\cite{llmcode,Copilot} have influenced both daily life and professional workflows.
\delete{Currently, leading AI companies, including OpenAI, Meta, and Google, release their LLMs as a service, allowing users to access their models by invoking the service, which is called \textbf{Language-Model-as-a-Service (LMaaS)}~\\~\cite{la2024language,sun2022black}.}
\add{These models are typically deployed under the \textbf{Language-Model-as-a-Service (LMaaS)} paradigm~\cite{la2024language,sun2022black}, where users send requests (\ie prompts) to a service endpoint and receive the generated responses.}

\add{The widespread adoption of LLMs and their integration into diverse software~\cite{weber2024large,shao2025llms,zhao2025llm} have led to LMaaS platforms processing millions of queries per day~\cite{qiao2024conserveharvestinggpuslowlatency}. Similar to traditional software services, the effective management of these platforms is critical for delivering a high-quality user experience, ensuring service availability~\cite{vayghan2021kubernetes}, and meeting Service Level Objectives (SLOs)~\cite{gambi2012modeling}.}. LMaaS management pursues two interdependent goals: (1) provisioning sufficient resources to match demand without over- or under-provisioning and (2) routing LLM requests to appropriate instances to maintain load balance and minimize latency. Failure to achieve either goal degrades service performance, potentially violating SLOs and resulting in economic losses or user attrition~\cite{OpenAIIncident}.

Although advancements in traditional service management via auto-scaling~\cite{karol2024self,meng2023deepscaler,luo2022power,yu2020microscaler,qiu2020firm} and request load balancing~\cite{rasmussen2008round,zhu2018improved,bhattacharya2024dynamically,cao2024queueflower} have enhanced management efficiency, the unique features of LMaaS reveal the limitations of these conventional approaches:
\noindent
\textbf{High workload variability exacerbates cold start issues}. As characterized in §~\ref{subsec:challenge1}, real-world LLM services exhibit substantial fluctuations in tokens per second (TPS). While weekday patterns emerge, daily peak TPS values vary unpredictably, rendering direct service workload predictions inherently imprecise. Traditional auto-scaling compensates for such forecast errors by reactively adjusting instance counts based on real-time metrics (\eg CPU utilization exceeding 80\%). While effective for conventional services with millisecond startup times, this reactive approach falters in LMaaS scenarios. The sheer scale of modern LLMs amplifies this issue. For instance, the Deepseek-R1 model~\cite{guo2025deepseek} with 671 billion parameters incurs cold start times ranging from tens of seconds. Such delays make reactive scaling impractical, as new instances can hardly be provisioned swiftly enough to address sudden demand spikes.
    
\noindent
\textbf{High request load variance intensifies load imbalance.} Unlike traditional cloud services, where requests of a given type exhibit consistent execution profiles, LLM request loads vary widely due to differences in response lengths. Our analysis of real-world LLM requests (§~\ref{subsec:challenge2}) reveals significant heterogeneity: noting that resource consumption, particularly GPU memory, scales with the response length, unlike the more stable resource usage of traditional requests. This dynamic and imbalanced execution load undermines conventional load-balancing strategies~\cite{rasmussen2008round,zhu2018improved}, such as round-robin and least-request approaches. Designed for uniform request profiles, these methods cannot adapt to the dynamic demands of LLM requests, leading to instance overloads.

These challenges underscore the necessity for management strategies tailored to the distinct dynamics of LLM services.
To meet this need, we introduce \nm, a hierarchical prediction-based LMaaS management framework aimed at optimizing resource utilization and reducing serving latency.
\nm consists of two main components: \emph{hierarchical prediction} and \emph{prediction-based management}.
The \emph{hierarchical prediction} component operates at two granularities. First, a \emph{service workload predictor} (§~\ref{sec:request_pattern_predictor}) leverages long-term patterns to forecast aggregate token demand. Second, a \emph{request load predictor} (§~\ref{sec:request_load_predictor}) estimates the individual request load in real time, informing a per-instance \emph{load anticipator} (§~\ref{sec:instance_load_anticipator}) of imminent load changes.
The \emph{prediction-based management} component uses these forecasts to optimize operations: An \emph{instance scaler} (§\ref{sec:proactive_instance_scaling}) proactively adjusts resources to mitigate cold-start issues, while a \emph{request router} (§\ref{sec:load-aware_request_router}) uses anticipated loads to balance instance loads, reducing tail latency.
This synergistic approach combines long-term workload trends with real-time request load dynamics, reducing dependency on reactive scaling, mitigating cold-start delays, and sustaining load equilibrium across LLM instances.

We evaluate \nm using open-source traces from Microsoft Azure LLM services~\cite{stojkovic2024dynamollm} and ShareGPT datasets~\cite{ShareGPT}. Our results demonstrate that, compared to state-of-the-art baselines, \nm reduces peak latency by over \delete{78.6\%}\add{45.1\%} while improving resource utilization by \delete{44.5\%}\add{49.38\%} in LMaaS systems under fluctuating workloads. Additionally, \nm effectively routes LLM requests to maintain load balance across instances, leading to a reduction of over \delete{59.1\%}\add{41.3\%} in P99 normalized latency and \delete{89.3\%}\add{66.58\%} in SLO violations. Notably, \nm introduces only a 0.23\% overhead relative to request latency, underscoring its practical effectiveness.

In summary, this paper presents the following key contributions: 
\begin{itemize}[leftmargin=*, topsep=2pt, parsep=0pt]
    \item Based on a detailed analysis of Azure LMaaS workloads, we identify two critical limitations of traditional service management techniques, creating opportunities for LMaaS solutions (§~\ref{sec:motivate}). 
    \item We introduce \nm, a novel hierarchical LMaaS management framework to combine long-term workload periodicity with short-term request load dynamics, optimizing resource utilization and reducing serving latency (§~\ref{sec:method}).
    \item Extensive evaluations indicate that \nm outperforms all baselines, significantly reducing serving latency and resource consumption, with low extra overhead (§~\ref{sec:evaluation}).
    We also open-source \nm at \cite{repo} to benefit both developers and researchers.
\end{itemize}

\section{Background}

\begin{figure}[t]
    \centering    
    \includegraphics[width=\columnwidth]{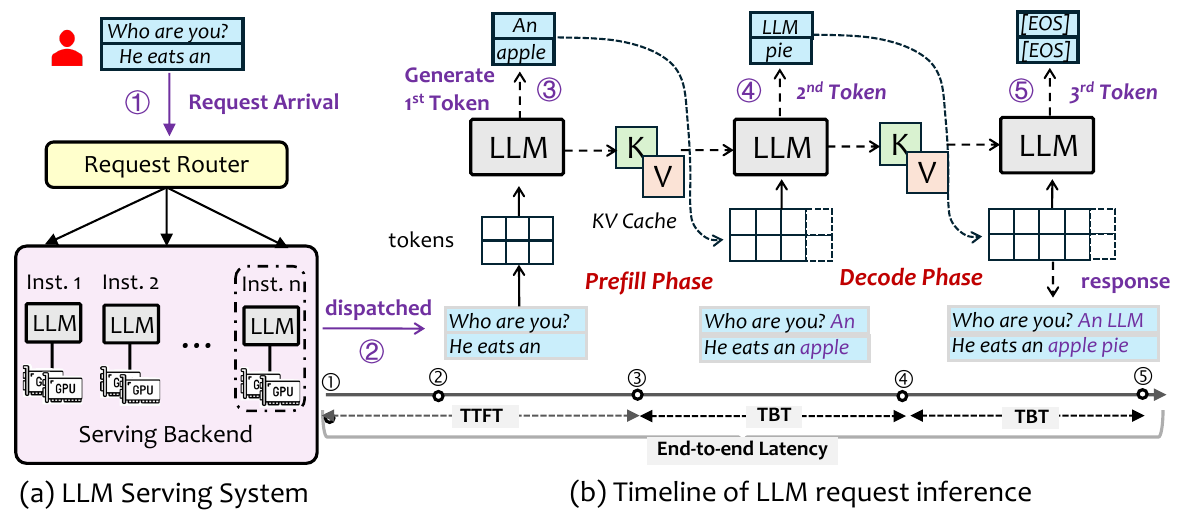}
    \caption{a) The framework of LLM serving system and b) an example of LLM request inference procedure.}
    \label{fig:background}
\end{figure}

A typical LMaaS system, illustrated in the left part of Fig.~\ref{fig:background}, consists of a request router and a serving back-end with multiple LLM instances.
The request router acts as the endpoint, receiving user requests and routing them to appropriate LLM instances within the serving back-end.
Each LLM instance hosts a complete model replica, supported by a group of dedicated hardware resources (\eg GPUs), enabling optimized inference through serving engines such as vLLM~\cite{kwon2023efficient}, SGLang~\cite{zheng2023efficiently}, and TensorRT-LLM~\cite{TensorRT}.
These engines typically employ advanced batching strategies (\eg continuous batching~\cite{yu2022orca}), which dynamically replace completed requests at each token step to maximize inference throughput.

\noindent
\textbf{Two-phase autoregressive generation.}
LLM services stream response tokens as they are generated.
As shown in the right part of Fig.~\ref{fig:background}, \delete{the inference process within each LLM instance for every LLM request} LLM inference is autoregressive: the model iteratively generates one response token based on the preceding tokens.
Specifically, the process begins with a single parallel computation over all input tokens in the prompt to produce the first token, known as the \emph{prefill phase}, which is compute-intensive\cite{zhong2024distserve}.
Subsequently, the inference enters the \emph{decode phase}, where the model generates one new token per iteration using the prompt and previously generated tokens until reaching an End-of-Sequence (EOS) token or a predefined limit. 
Each generated token is immediately sent to the request router and streamed to the user in real time.

\noindent
\textbf{KV Cache and memory management.}
During inference, intermediate key and value tensors from the attention mechanism~\cite{vaswani2017attention} are reused to generate subsequent tokens.
To avoid redundant computation, these tensors (KV cache~\cite{kwon2023efficient}) are stored in GPU memory, making the \emph{decode phase} highly memory-intensive~\cite{qin2024mooncake,patel2024splitwise}.
The KV cache size for each request grows with its token sequence length.
For instance, processing 64 batched Llama3-8B requests~\cite{dubey2024llama} with 8K tokens each can require up to 64 GB of KV cache, compared to 15 GB for model parameters. 
Given that modern GPUs offer only tens of GBs of memory (\eg 80 GB on NVIDIA H100), memory capacity becomes the primary bottleneck for LLM serving.
When GPU memory is exhausted, the inference engine must evict certain requests and free their KV cache; these requests are later reloaded, causing redundant recomputation and increased latency.

\noindent
\textbf{Metrics of LLM service quality.}
\label{sec:metrics}
Traditional services~\cite{gambi2012modeling,heidari2019cost,meng2023deepscaler} typically use end-to-end (E2E) latency as their Service Level Objective (SLO) metric
However, due to the dynamic execution and streaming nature of LMaaS, this metric alone is insufficient. 
Recent work~\cite{patel2024splitwise,zhong2024distserve} introduces alternative metrics better suited to LLM services:
1) \emph{Time to First Token (TTFT)}: the duration from the request submission to the receipt of the first generated token.
2) \emph{Time Between Tokens (TBT)}: the interval between sequentially generated tokens for each request.
3) \emph{Normalized E2E latency}: the E2E latency divided by the total number of output tokens.
These metrics collectively capture user experience: TTFT reflects responsiveness, TBT indicates streaming smoothness, and normalized E2E latency provides an overall efficiency measure.

\section{Motivation}\label{sec:motivate}

\subsection{Challenges in LMaaS Management}\label{sec:challenge}

To investigate LMaaS management challenges, we analyze open-source invocation traces from two Azure production LLM services, \emph{code} and \emph{chat}, spanning one week and 44.1 million requests~\cite{stojkovic2024dynamollm}. Each trace includes timestamps, prompt and response lengths. Unlike traditional request-based workloads, LLM workloads are inherently token-centric, as token throughput directly reflects computational demand. We further separate the analysis into prefill and decode phases due to their distinct resource characteristics.

\begin{figure*}[t]
    \centering
    \includegraphics[width=\textwidth]{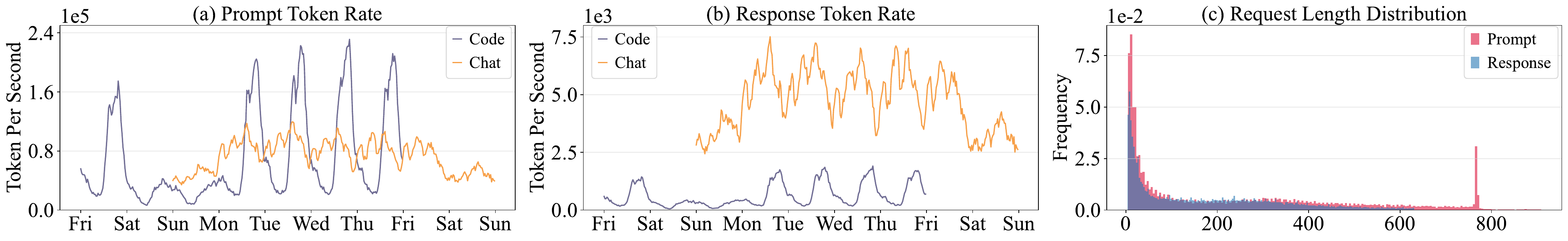}
    \caption{Azure LMaaS workload and ShareGPT datasets study: the distribution of the TPS for LLM (a) code and (b) chat service in Azure over one week, and (c) the prompt/response length in LLM conversations.}
    \label{fig:motivation}
\end{figure*}

\subsubsection{Challenge 1: Temporal Variability in LLM Service Workloads}\label{subsec:challenge1}
We first observe the workload variations across different LLM services from a service perspective. Fig.~\ref{fig:motivation}-(a) and (b) depict the average prompt and response tokens per second (TPS) for \emph{code} and \emph{chat} LLM services in Azure, aggregated over 30-minute intervals. Based on these trends, we identify three key observations:
\begin{enumerate}[leftmargin=*, topsep=0pt]
    \item \emph{LLM service workloads experience substantial fluctuations.} In the \emph{code} service, the peak prompt TPS exceeds its average by $3.3\, \times$ and its minimum by $35.6\, \times$, reflecting dramatic shifts within short timeframes. Such variability requires \delete{autoscaling mechanisms to dynamically adjust LLM instances, ensuring optimal resource utilization and maintaining service quality.}
    \add{dynamic resource scaling to avoid performance degradation and resource waste.}
    \item \emph{Distinct LLM services exhibit different periodic workload patterns.} Both services demonstrate a clear diurnal trend: peaking during working hours and dropping at night and on weekends, enabling proactive forecasting. However, notable differences exist between services; for example, the peak prompt TPS for the \emph{code} service are approximately twice those of the \emph{chat} service, while the \emph{chat} service's peak response TPS are nearly four times higher than those of the \emph{code} service. These differences stem from their distinct application domains, indicating that LMaaS management must be tailored to each service.
    \item \emph{Peak workload magnitudes introduce uncertainty.}
    Although weekday trends are consistent, daily peak TPS vary unpredictably (red lines in Fig.~\ref{fig:motivation}(a)–(b))
    For instance, the highest prompt TPS peak of \emph{code} service exceeds the lowest weekday peak by 35\%.
    This uncertainty hampers the accuracy of workload predictions, complicating effective resource management.

\end{enumerate}

\challenge{Service-specific provisioning heightens management intricacy, while extreme fluctuations and uncertain peak magnitudes strain the scalability and accuracy of existing management systems, thereby posing a risk of over-provisioning or degraded performance.}

\subsubsection{Challenge 2: Huge Variance of LLM Request Loads}~\label{subsec:challenge2}
In traditional cloud services, requests of a given type (\eg adding items or processing payments) exhibit similar execution profiles with minimal variability. Conversely, LLM service requests differ significantly due to variations in prompt complexity, necessitating a detailed analysis of inter-request differences.
To quantify this variability, we analyzed token counts of prompts and responses using ShareGPT datasets~\cite{ShareGPT}, which consists of over 90,000 real-world conversations collected from OpenAI’s chat service.
We present their [P5, P95] distributions to mitigate noise, as shown in Fig.~\ref{fig:motivation}-(c).

Our analysis reveals substantial variability in LLM request token counts.
Specifically, the prompt lengths vary significantly from 7 to 911 tokens, while responses range from 5 to 632 tokens, with median lengths of 52 and 87 tokens, respectively.
This broad spectrum reflects the diverse nature of user interactions, ranging from brief queries to elaborate, multi-sentence dialogues.
Such pronounced differences in prompt and response lengths translate directly into varying computational loads: longer prompts demand higher computation during the prefill phase, whereas extensive responses elevate memory usage and inference duration in the decode phase. 
Consequently, this variability undermines uniform request routing strategies, as heavy-load requests disproportionately consume resources, exacerbating load imbalances and potentially decreasing overall throughput, especially during peak usage periods.

\challenge{The wide range of prompt and response lengths result in highly variable request loads. This complicates the design of effective request routing policies, as the uncertainty of individual request loads increases the risk of load imbalance, leading to increased serving latency.}

\subsection{Opportunities in LMaaS Management}

The challenges of workload variability and diverse request loads highlight the limitations of traditional service management techniques and present opportunities for specialized LMaaS solutions.

\subsubsection{Opportunity 1: Hierarchical Service Workload and Request Load Prediction}
\label{sec:motivation_opportunity_1}

\begin{figure}[t]
    \centering
    \includegraphics[width=\columnwidth]{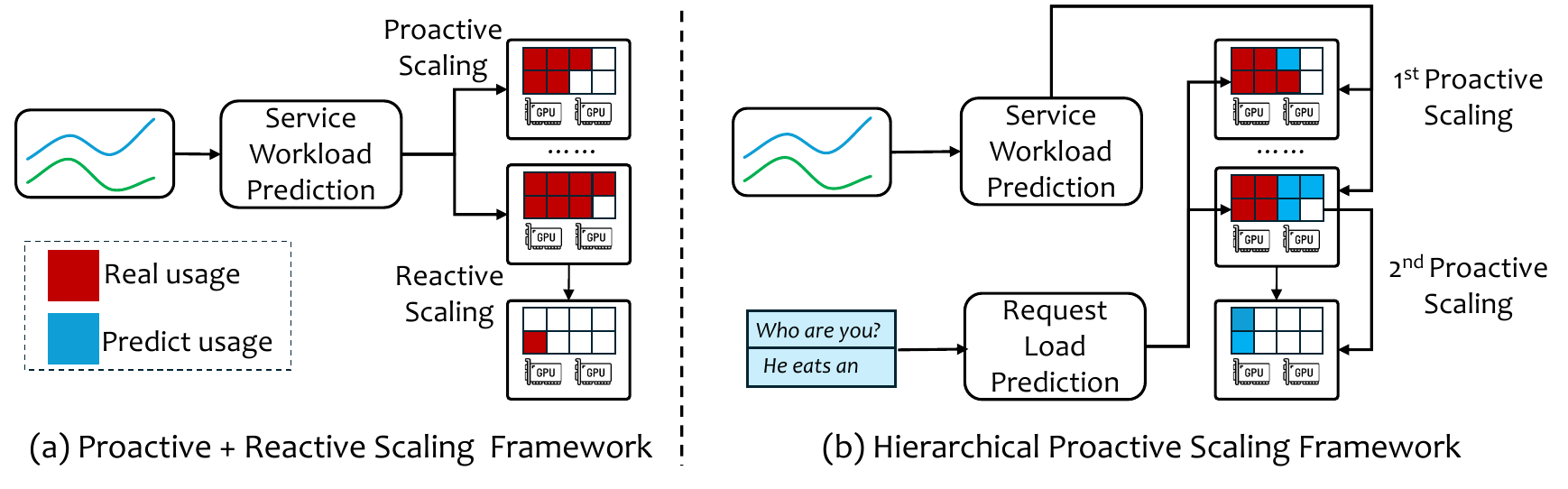}
    \caption{Different auto-scaling frameworks.}
    \label{fig:motivation_scaling}
\end{figure}

To cope with workload variability (Challenge 1), traditional services often use a hybrid scaling strategy (\eg Fig.~\ref{fig:motivation_scaling}-(a)): proactive scaling based on historical forecasts, supplemented by reactive scaling to handle prediction errors. This model, however, fails in the LMaaS context for two main reasons. First, the unpredictable workload peaks common in LMaaS lead to frequent forecast inaccuracies, necessitating reactive adjustments. However, the significant cold start latency of large LLM instances, often tens to hundreds of seconds~\cite{song2024funcscaler}, makes reactive scaling too slow to handle sudden demand spikes. Second, the high variance in per-request loads (Challenge 2) makes instance resource consumption highly volatile, triggering reactive scaling even more often and further exposing the impracticality of this approach.

These limitations highlight the need for a more proactive and adaptive strategy, naturally leading to a hierarchical approach combining service-level and request-level predictions (\ie Fig.~\ref{fig:motivation_scaling}-(b)):
\begin{enumerate}[leftmargin=*, topsep=0pt]
    \item \textbf{Service-level Workload Prediction:} Historical TPS patterns (\eg workday peaks) forecast aggregate demand, enabling instance scaling in advance for upcoming time windows.
    \item \textbf{Request-level Load Prediction:} Real-time inference data estimate the remaining token loads per request, allowing early resource adjustments to prevent instance overload.
\end{enumerate}
This hierarchical strategy addresses both long-term trends and short-term dynamics, reducing reliance on reactive scaling, mitigating cold start delays, and accommodating variable request loads. 

\opportunity{Designing a hierarchical prediction framework that integrates service-level workload with request-level load prediction could improve resource efficiency amid temporal variability and request heterogeneity.}

\begin{figure}[t]
    \centering
    \includegraphics[width=\columnwidth]{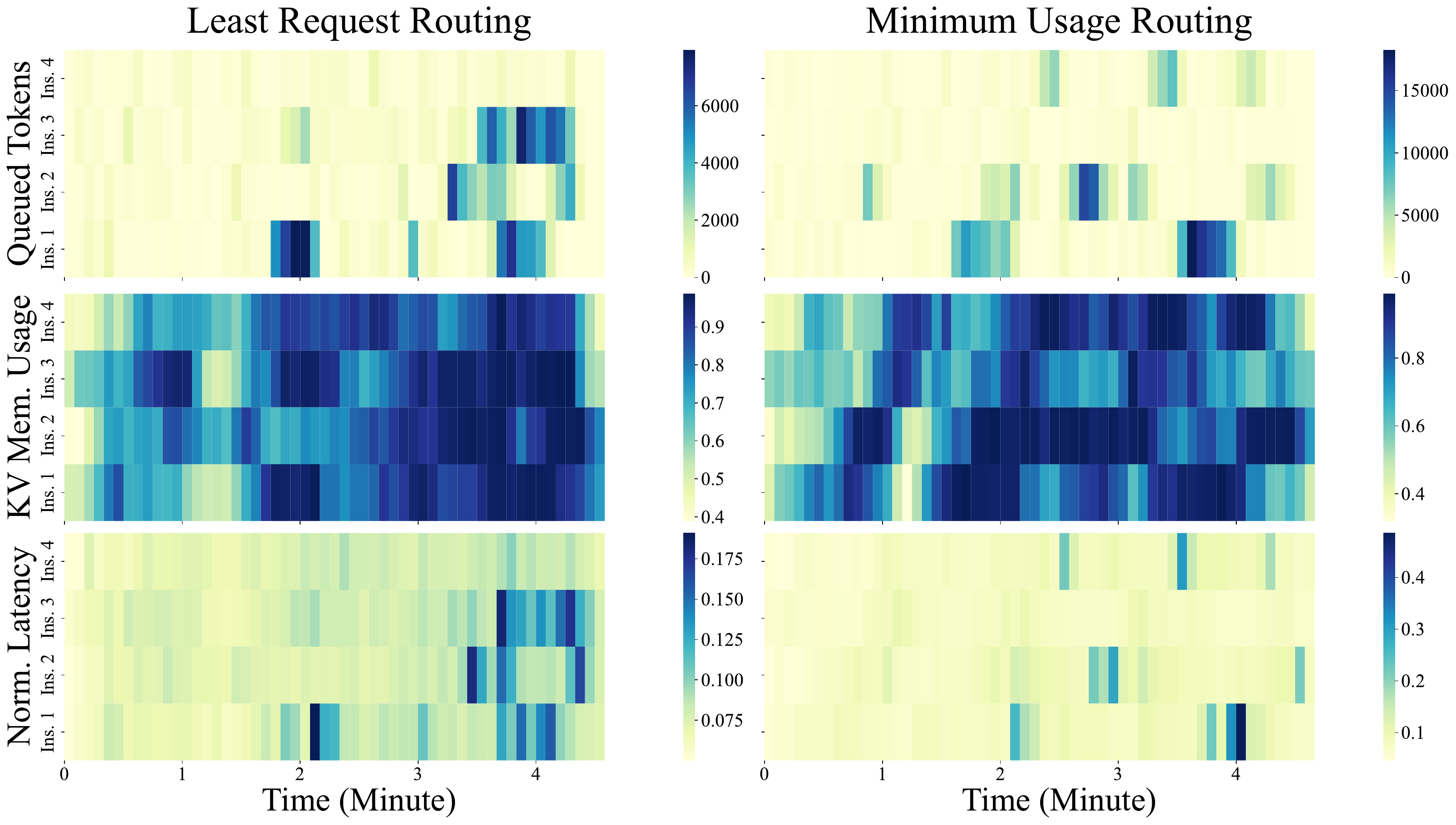}
    \caption{The performance timeline of two load-balancing algorithms on LMaaS benchmarks using four LLM instances.}
    \label{fig:motivation_load_balance}
    \vspace{-15pt}
    
\end{figure}

\subsubsection{Opportunity 2: Predicted Load-aware Request Routing.}
\label{sec:motivation_opportunity_2}
\delete{Effective request routing among a fixed set of instances is vital for maintaining LLM service quality, as instance overload can significantly impair key metrics like TTFT and normalized latency~\cite{cheng2024enabling,nie2024aladdin}. However, as highlighted in Challenge 2 (§~\ref{subsec:challenge2}), longer prompts require more computational resources in the prefill phase, while longer responses increase memory usage and processing time in the decode phase. Additionally, unlike traditional service requests, which typically exhibit relatively stable and statistical resource usage, the resource consumption of LLM requests, particularly GPU memory, scales linearly with response generation. Such imbalanced and dynamic execution loads render traditional load-balancing approaches~\cite{rasmussen2008round,zhu2018improved}, including round-robin, least-request, and minimum use, ineffective, as they assume uniform and static request loads.}

The high variance in request loads (Challenge 2) invalidates traditional load-balancing policies~\cite{rasmussen2008round,zhu2018improved}. Strategies like round-robin or least-request, which assume uniform request loads, can hardly handle the dynamic and heterogeneous resource demands of LLM inference, causing severe load imbalance and degraded service quality~\cite{cheng2024enabling,nie2024aladdin}.

To illustrate these shortcomings, we conducted a pilot benchmark that evaluated two common routing strategies: least request (LR) and minimum use (MU) in an LMaaS environment. We deployed four LLaMA2-7B instances~\cite{LlamaFamily} on a 4$\times$ A40 GPU cluster and simulated user traffic using ShareGPT prompts~\cite{ShareGPT} at a fixed queries-per-second (QPS) rate \add{(9.5 in this case, a value corresponding to the overload boundary that has been widely adopted in prior research~\cite{sun2024llumnix,qiu2024power,wu2024loongserve})}.
Three metrics were monitored: (1) \emph{Queued Request Tokens}, the queued prompt tokens awaiting GPU computation; (2) \emph{KV Memory Usage}, the percentage of GPU memory occupied by the KV cache, indicating decode-phase congestion; and (3) \emph{Normalized E2E Latency}, a comprehensive quality metric defined in Sec~\ref{sec:metrics}. The LR strategy routed requests to the instance with the fewest active requests, while the MU strategy targeted the instance with the lowest weighted average of GPU utilization and memory usage.
As shown in Fig.~\ref{fig:motivation_load_balance}, both approaches have serious limitations.
\begin{enumerate}[leftmargin=*, topsep=0pt]
    \item \textbf{LR:} At the 2-minute mark, instance 1 under LR exhibited a surge in prefill tokens, KV memory usage, and elevated normalized latency, signaling overload, while other instances remained underutilized. A similar imbalance recurred at 4 minutes. This stems from LR’s reliance on request counts alone, ignoring the significant request load variability.
    \item \textbf{MU:} Despite factoring in utilization, MU still encountered instance overloads, such as instance 2 at 3 minutes and instance 1 at 4 minutes, due to its reactive nature. When an instance reaches saturation, the ongoing requests continue to generate responses, progressively elevating resource demands and resulting in persistent instance overload.
\end{enumerate}
These findings underscore the critical need for routing strategies that explicitly account for LLM-specific load dynamics, moving beyond static or reactive paradigms.
An effective policy should proactively anticipate current and future instance loads to mitigate potential load imbalance.

\opportunity{LMaaS providers can leverage the request-level load prediction to anticipate each instance’s future request demands. This facilitates routing decisions that prevent instance overload, offering a proactive solution to the limitations of traditional load balancing.}

\section{Method}\label{sec:method}

\begin{figure}[t]
    \centering
    \includegraphics[width=\columnwidth]{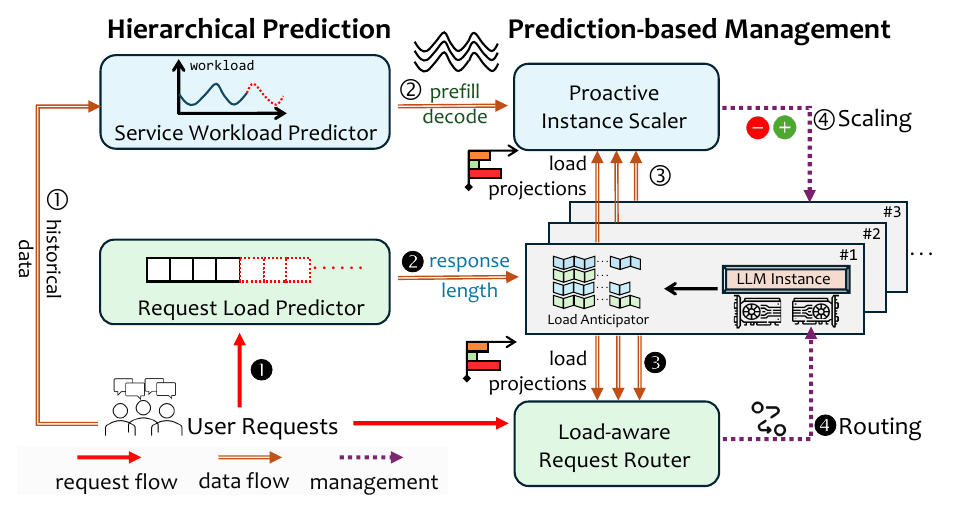}
    \caption{The overall framework of \nm.}
    \label{fig:framework}
    \vspace{-10pt}
    
\end{figure}

Motivated by our insights, we propose \nm, a hierarchical prediction-based LMaaS management framework, aiming to (1) dynamically autoscale LLM instances to optimize resource utilization under fluctuating workloads, and (2) predictively route LLM requests to achieve load balancing and reduce serving latency.

Fig.~\ref{fig:framework} overviews the workflow of \nm, comprising two main parts: \emph{hierarchical prediction} and \emph{prediction-based management}.
In long-term workload prediction, \nm forecasts the periodic token density for future time windows and determines the appropriate number of LLM instances required (§~\ref{sec:request_pattern_predictor}).
For short-term load prediction, \nm introduces a novel \emph{request load predictor} designed to estimate dynamic loads of individual LLM requests (§~\ref{sec:request_load_predictor}). For effective LMaaS management, the \emph{instance scaler} integrates hierarchical prediction results to optimize resource allocation, preventing resource over- or under-provisioning (§\ref{sec:proactive_instance_scaling}).
Simultaneously, the \emph{request router} considers both current and anticipated future loads across all instances when dispatching requests, ensuring instance load balancing and minimizing tail latency (§~\ref{sec:load-aware_request_router}).

\subsection{Tier-1: Service Workload Prediction}
\label{sec:request_pattern_predictor}

\begin{algorithm}[t]
\small
\caption{Offline Phase of Service Workload Prediction}
\label{alg:offline}
\LinesNumbered
\KwIn{
    Historical requests $\mathcal{R}$ and time windows $\mathcal{T} = \{T_1, T_2, \cdots\}$
}
\KwOut{
    Prediction model $\mathcal{M}_P, \mathcal{M}_D$ and profiled rate $\mu_p$, $\mu_d$, $\mu_{t}$
}

\For{$T_i \in \mathcal{T}$} {
    Aggregate the prefill and decode tokens within $T_{i-k}, \cdots, T_{i}$
    $\Rightarrow$ Generate $\mathbf{P} = \{P_{i-k}, \dots, P_i\}$, $\mathbf{D} = \{D_{i-k}, \dots, D_i\}$ \\ \label{algo:aggregation}
    
    $\mathbf{P} = \frac{\mathbf{P} - \min(\mathbf{P})}{\max(\mathbf{P}) - \min(\mathbf{P})}$, $\;$ $\mathbf{D} = \frac{\mathbf{D} - \min(\mathbf{D})}{\max(\mathbf{D}) - \min(\mathbf{D})}$ \tcp{Normalization} \label{algo:normalization}
    \tcc{Construct training datasets}
    Add training pair $\{P_{i-k}, \dots, P_{i-1}\} \rightarrow P_i$ to $S_P$ \\ \label{algo:train_p}
    Add training pair $\{D_{i-k}, \dots, D_{i-1}\} \rightarrow D_i$ to $S_D$ \\ \label{algo:train_d}

    \tcc{Update the maximum serving capability}
    \If{all requests in $T_i$ complete without an SLO violation} {
        $\mu_p = \max\left(\mu_p, \frac{\sum p(r) \in T_i }{|T_i|}\right)$, $\mu_d = \max\left(\mu_d, \frac{\sum d(r) \in T_i }{|T_i|}\right)$ \\
        $\mu_t = \max\left(\mu_t, \frac{\sum (p(r) + d(r)) \in T_i }{|T_i|}\right)$
    }
}
use $S_P$ to train $\mathcal{M}_P$ and use $S_D$ to train $\mathcal{M}_D$ \label{algo:train} 
\end{algorithm}

As illustrated in Sec.~\ref{subsec:challenge1}, LLM service workloads exhibit dynamism, along with clear temporal periodicity.
In light of this, \nm introduces a \emph{request workload predictor} to forecast long-term workload patterns and estimate the necessary number of LLM instances for upcoming coarse-grained time windows.
Prior research~\cite{qiu2020firm,luo2022power,zargarazad2023auto,meng2023deepscaler} has demonstrated the effectiveness of proactive auto-scaling strategies in optimizing the deployment of traditional services, including microservices and cloud services.

Nevertheless, the unique features of LLM services render simple prediction of request arrival patterns insufficient.
First, requests from different LLM services can exhibit varying prompt and response token distributions.
For instance, requests associated with conversational LLM services typically feature fewer prompt tokens and more response tokens compared to those for coding-related services.
Second, due to the two-phase generation process in LLM inference, both the prompt (\ie the compute-bound prefill stage) and the response (\ie the memory-bound decode stage) can become performance bottlenecks in the serving system~\cite{patel2024splitwise,qin2024mooncake}.
Consequently, we propose forecasting the arrival patterns of both prompt and response tokens for each individual LLM service.
This approach enables a more precise estimation of the number of LLM instances needed to handle future workload.

To accurately model and forecast token-level workload patterns of individual LLM services, we use the multiplicative Long Short-Term Memory (mLSTM) model~\cite{krause2016multiplicative}. The mLSTM integrates the strengths of Long Short-Term Memory (LSTM) networks~\cite{hochreiter1997long}, known for capturing both short-term and long-term patterns in time series data~\cite{li2019ea,sun2023sufs}, and multiplicative Recurrent Neural Networks (mRNN)~\cite{wu2016multiplicative}, which enhance traditional RNNs with multiplicative transition functions to model complex dependencies and interactions. By incorporating the intermediate state of the mRNN with the gating mechanisms in LSTM, mLSTM achieves efficient, accurate, and robust time series predictions from historical data.

In particular, \emph{service workload predictor} forecasts the prompt and response token densities for incoming requests within a fixed-length time window $T_i$, which is set to 10 minutes.
Moreover, to determine the required number of LLM instances to serve specific token density, it also profiles the serving capability of each LLM instance from historical data.
The application unfolds in two phases:

\noindent
\textbf{Offline training phase}.
During this phase, historical request data is collected for each LLM service over a brief period.
This data includes request timestamps, prompt tokens, and response tokens, which are used to train the mLSTM model.
The model takes as input the token counts from the previous $k$  time windows, denoted  $\{T_{i-k}, T_{i-k+1}, \dots, T_{i-1}\}$, and predicts the token count for the subsequent time window $T_i$.
To facilitate this, as shown in Alg.~\ref{alg:offline}, for each time window $T_i$, we aggregate the total number of prompt and response tokens within the previous $k$ windows (line  \ref{algo:aggregation}).
These collected pairs, \ie $\{P_{i-k}, \dots, P_{i-1}\} \rightarrow P_i$ and $\{D_{i-k}, \dots, D_{i-1}\} \rightarrow D_i$, are used to train the mLSTM models by learning the temporal patterns (line \ref{algo:train_p}, \ref{algo:train_d}, \ref{algo:train}). 
Additionally, the predictor also profiles the maximum throughput of prefill, decode, and total tokens of all historical time windows for each instance without any SLO violations, which represents the prefilling, decoding, and hybrid serving capability of one LLM instance of the specific LLM service.

\noindent
\textbf{Online prediction phase}.
When applied in the online phase, the model generates predictions for future time windows in a stepwise manner.
As illustrated in Alg.~\ref{alg:online}, at the beginning of each time window $T_i$ , the mLSTM model predicts the token counts  $P_{i+1}$  and  $D_{i+1}$  for the subsequent time window  $T_{i+1}$ .
This is accomplished through a two-step prediction process, where the model first predicts  $P_i$  and  $D_i$  for the current time window (line~\ref{algo:step1}), then uses those predictions to forecast  $P_{i+1}$  and  $D_{i+1}$ for next time window (line~\ref{algo:extend},~\ref{algo:step2}).
Based on the projections of both prompt and response tokens, the predictor calculates the required LLM instances in the next time window, \add{$N_{i+1}$} (line~\ref{algo:number}).
This look-ahead strategy facilitates the pre-initialization of LLM instances based on the projections, given that this initiation process typically requires dozens of seconds—well within the 10-minute time window.
Upon completion of the current time window~$T_i$, the ground truth values of  $P_i$  and  $D_i$  are appended to the input vectors  $\mathbf{P}$  and  $\mathbf{D}$, respectively, for the next prediction cycle (line \ref{algo:update}).
Additionally, the mLSTM model undergoes periodic re-training at longer intervals (\eg daily) using the most recent historical data.
This re-training ensures that the model remains adaptive to shifting request patterns and trends of LLM services.

\begin{algorithm}[t]
\small
\caption{Online Phase of Service Workload Prediction}
\label{alg:online}
\LinesNumbered
\KwIn{
    Historical aggregated token sequences: $\mathbf{P} = \{P_{i-k}, \cdots, P_{i-1}\}$, $\mathbf{D} = \{D_{i-k}, \cdots, D_{i-1}\}$
}
\KwOut{
    Estimated $N_1$ LLM instances for the next time window
}

\For{each current time window $T_i$} {
    Predict current window tokens: $\hat{P_i} = \mathcal{M}_P (\mathbf{P})$, $\hat{D_i} = \mathcal{M}_D (\mathbf{D})$  \\ \label{algo:step1}
    \tcc{Extend historical sequences}
    $\mathbf{P}' = \mathbf{P} + \{\hat{P_i}\}$, \quad $\mathbf{D}' = \mathbf{D} + \{\hat{D_i}\}$ \\ \label{algo:extend}

    Predict new window tokens: $\hat{P_{i+1}} = \mathcal{M}_P (\mathbf{P}')$, $\hat{D_{i+1}} = \mathcal{M}_D (\mathbf{D}')$ \\ \label{algo:step2}

    \tcc{Determine the required number of instances}
    $N_{i+1} = \max\left(\frac{\hat{P_{i+1}}}{\mu_p}, \frac{\hat{D_{i+1}}}{\mu_d}, \frac{\hat{P_{i+1}} + \hat{D_{i+1}}}{\mu_t}\right)$ \\ \label{algo:number}

    \If{$T_i$ has concluded} {
        Update historical sequences: $\mathbf{P} \gets \mathbf{P} + \{P_i\}$, $\mathbf{D} \gets \mathbf{D} + \{D_i\}$ \\ \label{algo:update}
    }
}
\end{algorithm}

\subsection{Tier-2: Request Load Prediction}
\label{sec:request_load_predictor}

While the workload predictor estimates the required number of LLM instances, prediction errors remain inevitable due to peak uncertainty and request burstiness~\cite{chen2016self,casale2011burn,wang2024burstgpt}.
Moreover, variability in LLM request loads complicates effective load balancing across instances (§~\ref{subsec:challenge2}).
\add{To mitigate these issues, \nm introduces request-level load prediction, which estimates the load of each individual request.
This fine-grained, dynamic awareness of each LLM request's load allows \nm to model and project the short-term load on each LLM instance.
Consequently, \nm can dynamically adjust LLM instances based on these immediate short-term load trends, rather than relying on historical data.
This approach also facilitates predictive, load-aware request routing.}

In detail, the load of each LLM request is influenced by both the prompt length, which is known beforehand, and the response length, which remains unknown until the request is completed.
Recent studies~\cite{zheng2024response,qiu2024power,jin2023s} have demonstrated that response lengths for specific LLM services exhibit a strong correlation with the semantic content of query prompts.
For example, translation requests typically yield response lengths comparable to the original prompt length.
Thus, predicting response length emerges as a viable strategy for estimating the load of individual LLM requests.
Nevertheless, designing such a predictor presents three main practical challenges:
\begin{enumerate}[leftmargin=*, topsep=0pt]
    \item \emph{Limited training data.} Compared to the broad input space of user prompts, only a small subset of LLM request data is available for training the predictor.
    \item \emph{Strict latency requirements.} Given the real-time nature and stringent latency demands of LLM services, the predictor must operate with minimal latency to ensure its applicability.
    \item \emph{Imbalance response length distribution.} As Fig.~\ref{fig:motivation}-(c) shows, LLM response lengths typically exhibit a long-tail distribution. Prior research~\cite{stocksieker2023data} has proven that such imbalanced data can introduce biases and complicate the development of a predictor that achieves consistently high accuracy across diverse requests.
\end{enumerate}

To tackle these challenges, we propose an effective and practical request load predictor, which is designed to take the prompt content as input and leverage its semantics to accurately predict the corresponding response length, \ie the number of decoded tokens.

In response to the limited training data, we select the pre-trained language model (PLM) as our proxy model, which is trained on large-scale corpora and has demonstrated strong semantic understanding capabilities~\cite{wang2023pre,min2023recent}.
We also employ prompt tuning~\cite{lester2021power} to adapt PLMs to the specific task of response length prediction.
This technique introduces a small set of additional parameters (\ie learnable prompt) into the model and optimizes these additional parameters rather than updating the entire model.
Compared to traditional tuning approaches~\cite{zheng2023learn,guo2019spottune}, prompt tuning can effectively mitigate the risk of over-fitting with limited training data, thereby enhancing the model’s performance in downstream tasks.

\begin{figure}[t]
    \centering
    \includegraphics[width=\columnwidth]{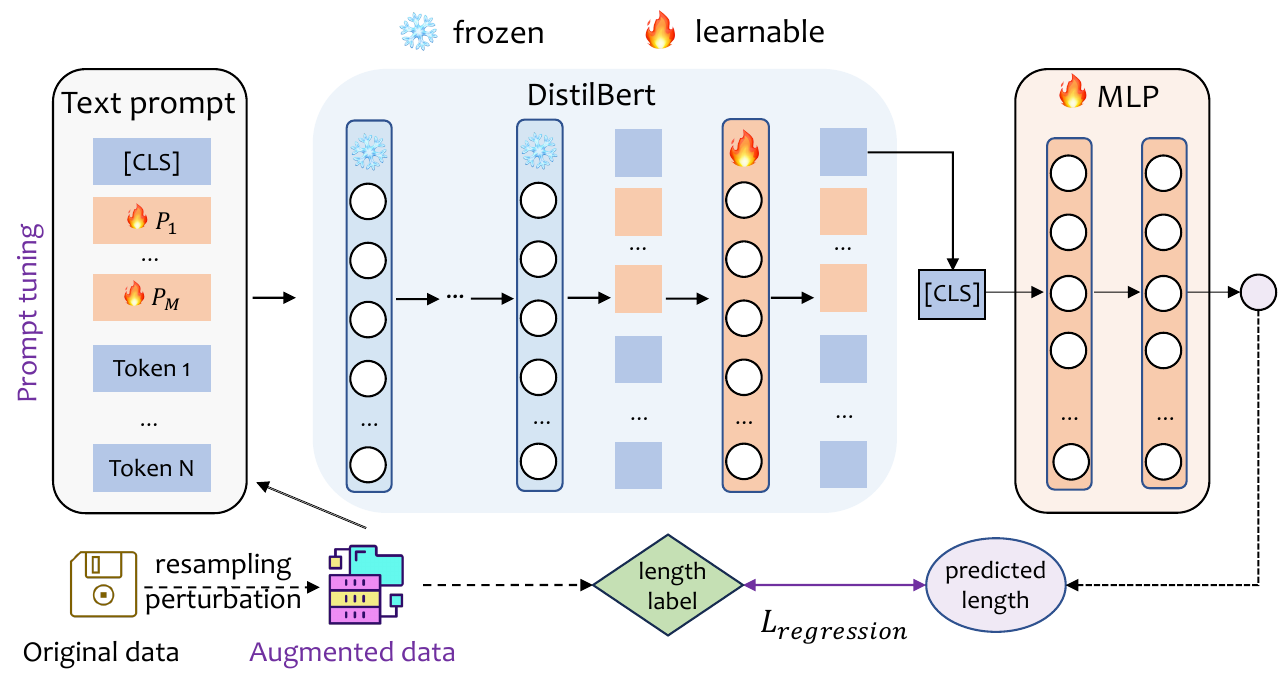}
    \caption{Request Load predictor training in \nm.}
    \label{fig:load_predictor}
\end{figure}

Specifically, given the stringent requirements for low overhead and latency, we select DistilBERT~\cite{sanh2019distilbert}, a distilled version of BERT~\cite{kenton2019bert}, which achieves a 60\% faster runtime while retaining over 95\% of BERT’s performance on previous benchmarks~\cite{wang2018glue}.
As illustrated in Fig.~\ref{fig:load_predictor}, the final hidden state of the first token (\ie [CLS]) from the DistilBERT output is extracted as it serves as the aggregate sequence representation for the subsequent regression task~\cite{kenton2019bert}.
This representation is passed through a two-layer feedforward neural network to generate the length prediction.
During the prompt-tuning procedure, we append $M$ additional learnable prompt tokens {$P_1, P_2, \dots, P_M$} to the input text tokens.
These prompt tokens are learnable and adaptively updated during fine-tuning.
To balance computational cost and performance, we also freeze the parameters of all DistilBERT layers except for the final layer.

To mitigate data imbalance issues, we adopt data augmentation techniques: \emph{over sampling}~\cite{estabrooks2004multiple} and \emph{text perturbation}~\cite{xu2020differentially}, to preprocess the original training data.
First, we partition the training samples into $N$ evenly spaced buckets based on response length, denoted as $\{B_1, B_2, \cdots, B_N\}$, where the bucket with the highest number of samples contains $S$ samples.
For buckets with fewer than $\mu S$ samples ($\mu=0.25$ in our experiments), we apply perturbation-based augmentation until the number of samples in each such bucket reaches $\mu S$.
Specifically, for each sample selected for augmentation, a small proportion (\eg 15\% ) of words in the prompt are randomly replaced with their synonyms while maintaining the same response length label.
Finally, the augmented dataset is used to train our request load predictor within a regression paradigm, consistently ensuring high accuracy for diverse LLM requests.

\subsection{Prediction-based Management}
Based on the hierarchical prediction results, \nm effectively manages instance scaling and request routing within LLM services.
Locally, each LLM instance maintains a \emph{load anticipator} that forecasts its load over an upcoming period based on request load prediction.
Centrally, the \emph{instance scaler} leverages both long-term workload forecasts and short-term instance load projections to proactively scale instances.
Concurrently, the \emph{request router} accounts for both the immediate and anticipated loads of individual instances to optimize load balancing and maintain high service quality.

\subsubsection{Instance Load Anticipator}
\label{sec:instance_load_anticipator}

\begin{figure}[t]
    \centering
    \includegraphics[width=\columnwidth]{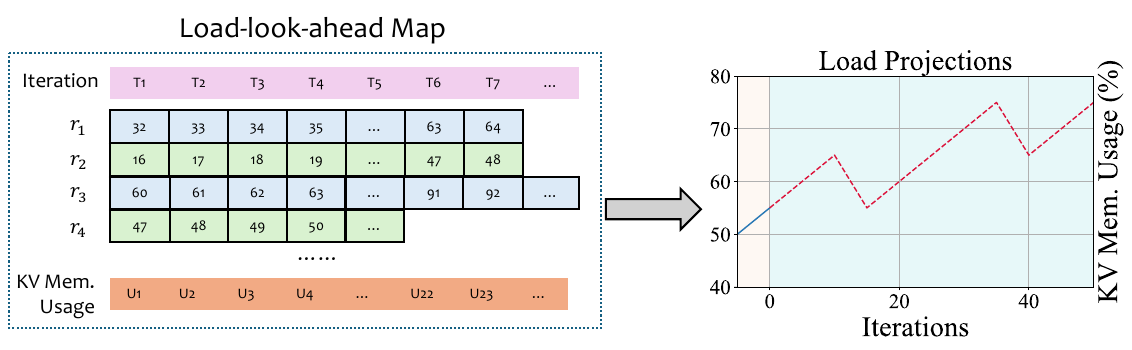}
    \caption{The load anticipator within each LLM instance.\delete{we update the $N_i$ to $U_i$}}
    \label{fig:lookahead_map}
\end{figure}

Based on the predicted request loads, the load anticipator integrates a \emph{load-look-ahead map} within each LLM instance, as illustrated on the left side of Fig.~\ref{fig:lookahead_map}.
This map estimates the KV cache memory usage for future iterations, motivated by the intuition that as the KV memory approaches saturation, preemption of requests will occur, which in turn affects the serving latency of LLM requests.
Specifically, the load-look-ahead map records $\mathbf{U}$, the fraction of total KV tokens relative to total token capacity $M$, for the next $L$ iterations, where $L$ is the maximum output tokens of the LLM (\eg 4096 tokens for LLaMA-2).
For a new LLM request $r_n$, with $P$ prompt tokens and $D$ predicted response tokens, the map is updated once the prefill phase is complete.
This involves adjusting the map as follows:   $U_i^{'} = \frac{U_i * M + (P + i)}{M}, i \in [0, D)$.

To mitigate inaccuracies in the predicted response length, where the ground truth response length $\hat{D}$ may differ from $D$, we refine the map to ensure more accurate projections.
In detail, if a request completes earlier than predicted (\ie $\hat{D} < D$), the estimated token consumption for the remaining iterations is subtracted: $U_i^{'} = \frac{U_i * M - (P + i)}{M}, i \in [\hat{D}, D)$.
Conversely, if a request does not complete within the predicted number of iterations (\ie $\hat{D} > D$), we introduce a ``virtual'' extension of the response length (set to 0.2$\hat{D}$ in our implementation) and update the map accordingly.
This process is repeated iteratively until the request completes, with any exceeded iterations being subtracted in the same manner.

Using the constructed load-look-ahead map, the load anticipator can project the instance’s load over the next $l$ iterations (\eg $l = 100$), as shown on the right side of Fig.~\ref{fig:lookahead_map}, which serves as a guide for both instance scaling (§~\ref{sec:proactive_instance_scaling}) and request routing (§~\ref{sec:load-aware_request_router}).

\subsubsection{Proactive Instance Scaler}
\label{sec:proactive_instance_scaling}

As discussed in Sec.~\ref{sec:motivation_opportunity_1}, to mitigate workload prediction inaccuracies and the prolonged cold-start delays associated with LLM services, \nm proposes a hierarchical prediction-based mechanism for instance scaling.
Specifically, this mechanism (1) periodically estimates the required number of instances for upcoming time windows based on long-term workload patterns and (2) continuously tracks short-term load projections of instances, dynamically adjusting the instance count accordingly.
This two-tiered approach can effectively prevent both the over-provisioning and under-provisioning of LLM resources.

At the beginning of each prediction time window $T_i$, if the predicted instance count $N_{i+1}$ exceeds the current count $N_c$, additional $N_{i+1} - N_c$ LLM instances are pre-initialized.
Conversely, if $N_{i+1} < N_c$, the instance scaler conservatively scales down by isolating $N_c - N_{i+1}$ instances from the request routing process, allowing their ongoing requests to complete naturally.

During each time window, the instance scaler further adjusts the number of instances following a ``one potentially overloaded instance, one additional instance scaled up'' policy.
An instance is regarded as potentially overloaded if, as predicted by its load anticipator for the next $l$ iterations, its KV memory usage exceeds 95\% in more than 10\% of those iterations.
This proactive adjustment mechanism enables earlier detection of overloads compared to purely reactive methods, thereby predictively scaling up to mitigate SLO violations.
Moreover, to prevent excessive resource provisioning, a scaling-down action is activated if the predicted memory utilization across all instances remains consistently below a specified threshold, $T_f$ (set at 30\%), throughout the next $l$ iterations.
This scaling-down action will only be triggered once within each time window to avoid thrashing~\cite{zhang2022adaptive}, with the number of isolated instances being $N_c - \frac{\sum_{ins.}\max (U^{'})}{T_f}$.

\subsubsection{Load-aware Request Router}
\label{sec:load-aware_request_router}

The \emph{request router} in \nm is designed to distribute LLM requests across multiple instances to mitigate instance overload and enhance serving efficiency by taking into account both current and anticipated future loads.
Incoming requests are first subject to a queuing mechanism: if all instances are overloaded, the request is queued. If the queue capacity is surpassed, the request is aborted, and the user is notified.
For each incoming request $r$ with $P$ prompt tokens and $D$ response tokens to be predicted, the router estimates the potential load for each instance if request $r$ is assigned to it.
Specifically, the estimated load of i-th instance comprises three components: prefill load $L_{p_i}$, decode load $L_{d_i}$, and memory overflow risk $L_{m_i}$ due to request evolving.
The first two terms are computed as $L_{p_i} = \text{queued\_}p_i+ P\; $ and $\; L_{d_i} = \text{current\_}d_i + D$, respectively, where $\text{queued\_}p_i$ represents total number of current queued token to be prefilled and $\text{current\_}d_i$ represent the total tokens to be generate in the decode phase.
These two metrics reflect the immediate computational and memory demands on each instance.

To further mitigate the risk of KV memory overflow, the router queries each instance's load anticipator.
The load anticipator ``virtually'' adds request $r$ into its load-look-ahead map and returns the peak memory usage $U_k$ anticipated over the next $l$ iterations.
Based on this projected usage, the router calculates a memory penalty component for the $i$-th instance as: $L_{m_i} = \max (0,\; U_k - T_{mem}) * M,$,
where $T_{mem}$ is the ideal memory usage that would not incur preemption latency (set to 80\% in our experiments).

Finally, the request router dispatches $r$ to the instance with the minimum estimated load by solving: $\mathop{\arg\min}\limits_{i \in [1, N]} L_i = L_{p_i} + L_{d_i} + \beta \times L_{m_i}$,
where $\beta$ is the penalty term coefficient, which is set to 1.

\section{Evaluation}
\label{sec:evaluation}

In this section, we conduct experiments to evaluate \nm by answering the following research questions (RQs):

\begin{itemize}[leftmargin=*, topsep=0pt, parsep=0pt]
    \item \textbf{RQ1:} How accurate is \nm in the hierarchical prediction?
    \item \textbf{RQ2:} How effective is \nm in LLM instance scaling?
    \item \textbf{RQ3:} How effective is \nm in LLM request routing?
    \item \textbf{RQ4:} What extra overhead does \nm incur in  management?
\end{itemize}

\subsection{Evaluation Setup}

\noindent
\textbf{Implementation.}
\nm is implemented in 3.5K lines of Python code and designed in a non-intrusive paradigm, allowing seamless adaption to different serving frameworks and LLMs.
In our experiments, we utilize the popular open-source LLM family, LLaMA~\cite{LlamaFamily}, as our back-end LLMs.
For the LLM inference engine \add{within each LLM instance}, we employ vLLM~\cite{vllm} (version 0.6.6), one of the most commonly used LLM serving frameworks~\cite{kwon2023efficient}.

\noindent
\textbf{Testbed.}
We conduct our experiments on an Ubuntu 22.04.1 GPU server configured with a 96-core Intel Xeon Gold 6342 CPU running at 2.80 GHz, 512 GB of RAM, and eight NVIDIA A40 GPUs, each with 40 GB of memory, connected via PCIe.

\noindent
\textbf{Workloads.}
To emulate real-world LLM service invocations, we use the Azure LLM inference trace 2024~\cite{stojkovic2024dynamollm}, which records 44 million requests from two production services, \textit{azure-code} and \textit{azure-chat}, over one week.
Each trace entry includes timestamps and token counts for both prompts and responses, allowing us to reproduce request arrivals and enforce exact token generation by setting \textit{ignore\_eos} to true and \textit{max\_tokens} to the desired response length, following prior benchmarks~\cite{kwon2023efficient,lazuka2024llm,qiu2024power,wang2024burstgpt}.
Since these traces omit prompt content, we further use the ShareGPT dataset~\cite{ShareGPT}, which provides over 90,000 real-world LLM conversations with detailed prompts and responses, to evaluate \nm’s request router that predicts response lengths from prompt semantics.

\noindent
\textbf{Metrics.}
Our evaluation employs key metrics as introduced in Sec.~\ref{sec:metrics} for assessing LLM service quality.
In detail, we present results for \emph{Time-to-First-Token (TTFT)} and \emph{Normalized Latency}, which respectively represent the first token latency and the average latency per token.
Additionally, we quantify Service Level Objective (SLO) attainment, defined as the proportion of requests fulfilled within the predetermined SLO threshold.
Following prior studies~\cite{patel2024splitwise}, we set the SLO for Norm. Latency.
It is set at 3$\times$ the \add{normalized} latency measured for a single, isolated request execution \delete{within the system, \ie 0.2s} \add{under no system contention}.
\add{This value is determined by averaging the execution times of five independent runs, with the final SLO value being 0.2s.}

\subsection{RQ1: Accuracy of Hierarchical Prediction}

\begin{table}[tbp]
    \centering
    \caption{\delete{The updated table.} Mean and maximum absolute percentage error (APE) of workload prediction in Azure datasets (10min-window).}
    \vspace{-10pt}
    \label{tab:RQ1_1-workload_predictor}
    \resizebox{\linewidth}{!}{%
        \begin{NiceTabular}{l||cccc|cccc}
        \CodeBefore
        \Body
            \toprule
                 \multirow{3}{*}{Methods} & \multicolumn{4}{c}{\add{Mean APE}} & \multicolumn{4}{c}{\add{Max APE}} \\
                \cmidrule{2-9}
                 & \multicolumn{2}{c}{Azure-code} & \multicolumn{2}{c}{Azure-chat} & \multicolumn{2}{c}{Azure-code} & \multicolumn{2}{c}{Azure-chat} \\
                 & prompt & response & prompt & response & prompt & response & prompt & response \\
                \midrule
                ARIMA & 59.17\% & 61.44\% & 15.94\% & 16.12\% & 91.00\% & 90.18\% & 74.03\% & 82.09\% \\
                ETS & 54.63\% & 55.55\% & 15.93\% & 16.10\% & 86.71\% & 83.54\% & 73.95\% & 81.96\% \\
                Prophet & 26.26\% & 28.49\% & 8.05\% & 8.28\% & 67.88\% & 62.27\% & 27.30\% & 25.12\% \\
            \midrule
            \multirow{3}{*}{\makecell{~\\ \nm ~\\ average}} & \textbf{7.74\%} & \textbf{8.45\%} & \textbf{4.15\%} & \textbf{4.30\%} & \textbf{26.25\%} & \textbf{30.30\%} & \textbf{21.16\%} & \textbf{19.88\%} \\
            \cmidrule(lr){2-3}  \cmidrule(lr){4-5} \cmidrule(lr){6-7} \cmidrule(lr){8-9} 
              & \multicolumn{2}{c}{\textbf{8.10\%}} & \multicolumn{2}{c|}{\textbf{4.23\%}} & \multicolumn{2}{c}{\textbf{28.28\%}} & \multicolumn{2}{c}{\textbf{20.52\%}} \\
              \cmidrule(lr){2-5} \cmidrule(lr){6-9}
              & \multicolumn{4}{c}{\textbf{6.17\%}} & \multicolumn{4}{c}{\textbf{24.40\%}} \\
            \bottomrule
        \end{NiceTabular}%
    }
     \vspace{-15pt}
\end{table}

\subsubsection{Accuracy of Service Workload Predictor}
We first assess the accuracy of the workload predictor in \nm using Azure traces from \emph{Azure-code} and \emph{Azure-chat} services.
To conduct this evaluation, we split the original dataset into a training set and a test set in a 1:1 ratio based on chronological order, ensuring that the former is used for model training while the latter is used for accuracy evaluation.
\add{The time window size is set to 10 minutes, which is slightly longer than the common scaling-up time of LLM instances (e.g., several minutes).}
For the evaluation metric, we adopt absolute percentage error (APE), defined as $\frac{|\text{prediction} - \text{actual}|}{\text{actual}}$.  We report both the mean and maximum APE values across all time windows.

\noindent
\textbf{Baselines.}
We select several widely-used time series forecasting algorithms: \textit{autoregressive integrated moving average} (ARIMA)~\cite{box1970distribution}, \textit{exponential smoothing} (ETS)~\cite{gardner1985exponential}, and Prophet~\cite{taylor2018forecasting}, for comparison.
\add{All these models are trained only using CPUs. }

\begin{figure*}[t]
    \centering
    \includegraphics[width=\textwidth]{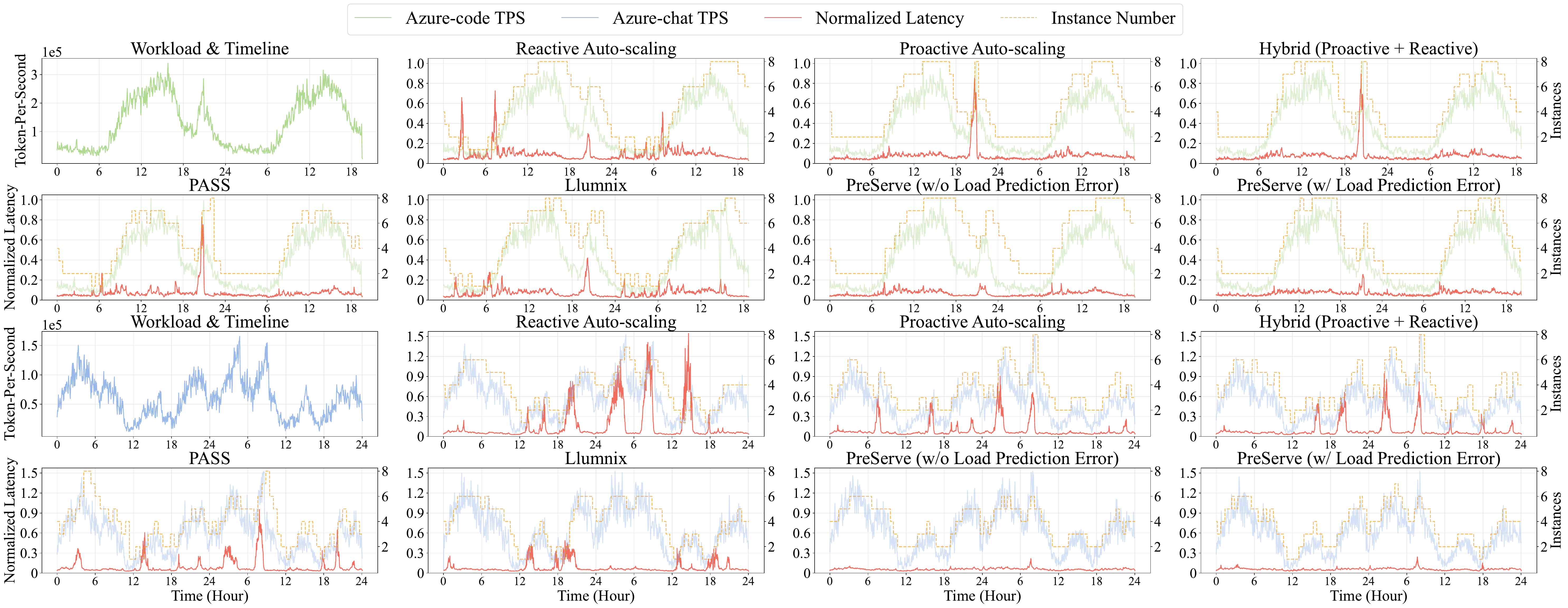}
    \caption{The evaluation results aggregated over 5-minute intervals under real-world workloads, using various scaling strategies.}
    \label{fig:RQ2_instance_scaling}
\end{figure*}

\noindent
\add{\textbf{Results.}}
The accuracies are presented in Tab.~\ref{tab:RQ1_1-workload_predictor}, demonstrating that \nm consistently outperforms all baseline methods. Specifically, \nm achieves an average mean APE of 8.10\% and 4.23\% for the code and chat services, surpassing the second-best baseline, Prophet, by 70.4\% and 48.2\%, respectively.
This confirms the effectiveness of \nm in effectively capturing periodic workload patterns of LLM services.
Nevertheless, we also observe that all methods exhibit high maximum APE values, \eg ARIMA reaches a maximum APE of 91\% in predicting prompt tokens for code service.
Despite \nm's superior performance, it still presents an average maximum APE of 24.4\% for individual time windows.
Such discrepancies can potentially lead to significant resource over- or under-provisioning, emphasizing the necessity of implementing hierarchical proactive scaling strategies.

\begin{table}[tbp]
    \centering
    \caption{The response length prediction accuracy (up to 4096).}
        \vspace{-10pt}
    \label{tab:RQ2_1-load_predictor}
    \resizebox{0.88\linewidth}{!}{%
        \begin{NiceTabular}{l||cccc}
        \CodeBefore
        \Body
            \toprule
                \rowcolor{gray!15} Methods &  MAE & Acc-25 & Acc-50 & Acc-100 \\
                \midrule
                $\mu$-Serve & 355.59 & 32.31\% & 49.35\%  & 65.25\% \\
                PiA (Vicuna-13B) & 283.86 &  39.27\% & 54.18\% & 68.56\% \\
                PiA (ChatGPT) & 127.41 &  50.42\% & 61.25\% & 70.34\%  \\
            \midrule
            \nm & \textbf{78.25} & \textbf{56.77\%} & \textbf{68.79\%}  & \textbf{77.95\%} \\
            \textit{Improvement} & $\uparrow$ (38.6\%) & $\uparrow$ (12.6\%) & $\uparrow$ (12.3\%) & $\uparrow$ (10.8\%) \\
            \bottomrule
             \end{NiceTabular}%
     }
     \vspace{-16pt}
\end{table}

\subsubsection{Accuracy of Request Load Predictor.} 
We quantitatively evaluate the accuracy of the request load predictor in estimating response lengths using the ShareGPT dataset, which is randomly divided into training and testing sets in a 7:3 ratio.
The evaluation employs the mean absolute error (MAE) metric to measure the average difference between the predicted and actual number of tokens.
Additionally, we consider three accuracy thresholds: \emph{Acc-25}, \emph{Acc-50}, and \emph{Acc-100}, where a prediction is deemed correct if the difference is within 25, 50, or 100 tokens, respectively.

\noindent
\textbf{Baselines.}
We compare the request load predictor in \nm with two state-of-the-art (SOTA) baselines: \emph{$\mu$-Serve} and \emph{PiA}.
$\mu$-Serve~\cite{qiu2024power} reformulates response length prediction as a classification task by dividing response lengths into multiple buckets.
It fine-tunes BERT and a multilayer perceptron to classify LLM requests into these predefined length categories.
We adopt the original implementation of $\mu$-serve and use the median value of each bucket as the predicted length for its respective class.
PiA (Perception in Advance)\cite{zheng2024response} modifies the original prompt by incorporating an instruction to estimate the response length in advance.
For comparisons, we utilize two base LLMs: Vicuna-13B-v1.5\cite{Vicuna} and ChatGPT~\cite{ChatGPT}.

\noindent
\add{\textbf{Results.}}
Tab.~\ref{tab:RQ2_1-load_predictor} presents the accuracy results.
Despite employing the smaller DistilBERT model, \nm significantly outperforms the BERT-based $\mu$-Serve across all metrics, with a 78.0\% improvement in MAE and a 19.5\% improvement in Acc-100.
These advances demonstrate the effectiveness of the training procedure in \nm.
Furthermore, \nm also surpasses PiA when used with both Vicuna-13B and ChatGPT, achieving increases of 38.6\% and 12.3\% in MAE and Acc-50, respectively.
The inferior performance of PiA stems primarily from its lack of fine-tuning to adapt the model from general language capabilities to the downstream response-length prediction task.
In addition, PiA requires intrusive modifications to prompts and LLMs during inference, introducing substantial overheads that restrict its practical applicability.
In contrast, \nm optimizes the fine-tuning of a smaller, distilled model to balance the overhead and accuracy of response-length prediction.

\subsection{RQ2: Effectiveness for Instance Scaling}

\begin{figure*}[t]
    \centering
    \includegraphics[width=\textwidth]{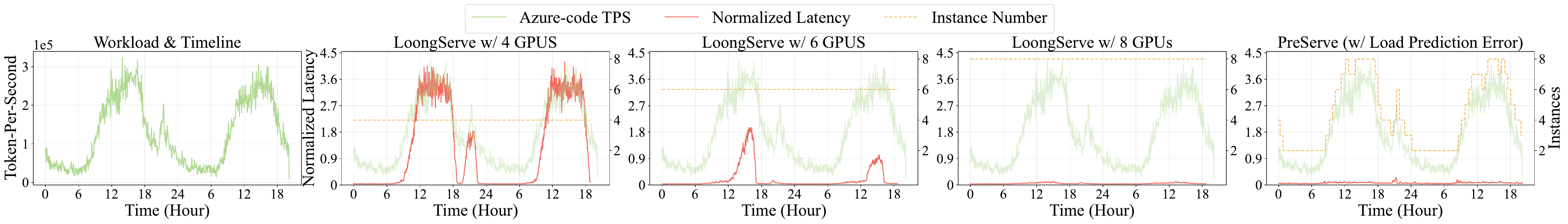}
    \caption{Evaluation of LoongServe compared to \nm on the Azure-code workload across various resource configurations.}
    \label{fig:RQ2_LoongServe}
\end{figure*}

\noindent
\add{\textbf{Settings.}}
In this RQ, we evaluate the effectiveness of \nm's hierarchical proactive scaling strategy in ensuring service quality and optimizing resource efficiency.
For experimental workloads, we utilize real-world traces from the \emph{Azure-code} and \emph{Azure-chat} services~\cite{stojkovic2024dynamollm}.
\add{However,} these publicly available traces typically omit prompt contents due to privacy concerns.
\add{To isolate the assessment of scaling capabilities,} we \add{first} directly employ ground-truth response lengths for the request load predictor to independently assess scaling capabilities, \add{a configuration we denote as \nm w/o load errors.}
\add{Besides, to simulate a more realistic scenario, we also introduce prediction errors based on the error proportion distribution from RQ1 Tab.~\ref{tab:RQ2_1-load_predictor}, which we term \nm w/ load errors.}
We use data from the first two days to train the workload predictor and the subsequent two days' data to generate the workload.
\add{Lastly}, we deploy up to eight LLaMA-2-7B instances and allow algorithms to dynamically auto-scale instances.

\noindent
\textbf{Baselines.}
We compare \nm against \add{five} service-level management baselines:
(1) \emph{Reactive}: it leverages the current instance status (\ie KV memory usage) to reactively scale LLM instances up or down.
(2) \emph{Proactive}: it uses predicted LLM workloads to proactively determine the required number of LLM instances.
(3) \emph{Hybrid (proactive + reactive)}: it combines workload predictions and current instance status to dynamically determine the appropriate number of LLM instances.
\add{(4) \emph{PASS}~\cite{guo2024pass}: it combines workload forecasting with a serving performance model to proactively scale instances, aiming to minimize resource costs while meeting strict SLOs. For our comparison, we adopt its scaling strategy into the LMaaS instance scaler following the implementation described in the original paper.}
\add{(5) \emph{Llumnix}~\cite{sun2024llumnix}: it introduces the concept of ``virtual memory'' for each LLM request and dynamically reschedules ongoing requests to prevent instance overload, as well as performs advance auto-scaling of instances. We use its open-source implementation~\cite{LLumnix-artifact} for our comparison.}
\add{Furthermore, to distinguish our service-level approach from instance-level optimizations, we also compare \nm against LoongServe~\cite{wu2024loongserve}, which dynamically adjusts the degree of parallelism for each request at runtime to improve serving efficiency.
Since LoongServe is designed to operate on a fixed set of resources, we evaluate its performance on dynamic LMaaS workloads across different GPU allocations, using its open-source implementation~\cite{LoongServe-artifact}.}

\noindent
\add{\textbf{Results of Service-level Management Comparison.}}
We derive the following key observations based on Fig.~\ref{fig:RQ2_instance_scaling}:
\begin{itemize}[leftmargin=*, topsep=0pt, parsep=0pt]
    \item Reactive auto-scaling frequently introduces local latency spikes due to its untimely scaling response, primarily caused by the cold start time issue of LLM instances.
    \item Proactive, hybrid, and PASS, all fail to allocate instances in advance for unexpected workload spikes, \eg around 20 hours in the code service workload. In detail, PASS causes a high average peak normalized latency of 0.89s. This demonstrates that these approaches lack the capability to handle prediction errors.
    \item While Llumnix is the best-performing baseline, its ``virtual memory'' mechanism and runtime request rescheduling still struggle with sudden workload surges. Consequently, it yields a high average peak latency of 0.455s. Furthermore, we observed frequent and unstable spikes in normalized request latency with Llumnix, which we attribute to the communication overhead and delays from KV cache migration during its runtime rescheduling.
    \item \nm w/o load error achieves the best performance, maintaining a low average peak normalized latency of 0.197s. However, even when integrating the load error, \nm still achieves superior performance, maintaining the average peak normalized latency within 0.250s. This robustness stems from two key factors: the high prediction accuracy of \nm's load predictor and the error refinement mechanism within the load anticipator.
\end{itemize}
In summary, \nm's hierarchical proactive strategy enables it to pre-allocate instances based on predicted workload patterns while adapting to unforeseen request spikes by anticipating instance loads. Compared to Llumnix, the top-performing baseline, \nm reduces the peak normalized latency by an average of 45.1\% to 0.250 seconds.
Furthermore, when compared against a static eight-instance allocation, \nm reduces resource consumption by an average of 49.38\% with minimal SLO violations, still outperforming the SOTA Llumnix (which achieves a 48.26\% reduction but suffers significant SLO violations). These results demonstrate \nm's superior ability to enhance service quality and resource efficiency.

\noindent
\add{\textbf{Results of Instance-level Inference Framework Optimization Comparison.}
Fig.~\ref{fig:RQ2_LoongServe} presents a performance comparison between \nm, and LoongServe, a framework focused on instance-level optimization. With a static allocation of 4 GPUs, LoongServe's normalized latency reached 4.20s, significantly violating SLOs. Even with 6 GPUs, it could not accommodate workload peaks, leading to a peak latency of 1.96s. Although LoongServe's performance with 8 GPUs became comparable to that of \nm, our system achieved this using only 4.35 GPUs on average, demonstrating superior resource efficiency.
These results highlight that instance-level optimizations, which operate within a fixed resource paradigm, are inherently insufficient for the highly dynamic workloads of LMaaS.
In contrast, \nm focuses on service-level management to intelligently scale instances and route requests, which is essential for maintaining performance under fluctuating workloads.
Furthermore, as \nm's design is orthogonal to these instance-level approaches, they can also be integrated into the managed instances of \nm for further gains.}

\subsection{RQ3: Effectiveness for Request Scheduling}

\begin{figure*}[t]
    \centering
    \includegraphics[width=\textwidth]{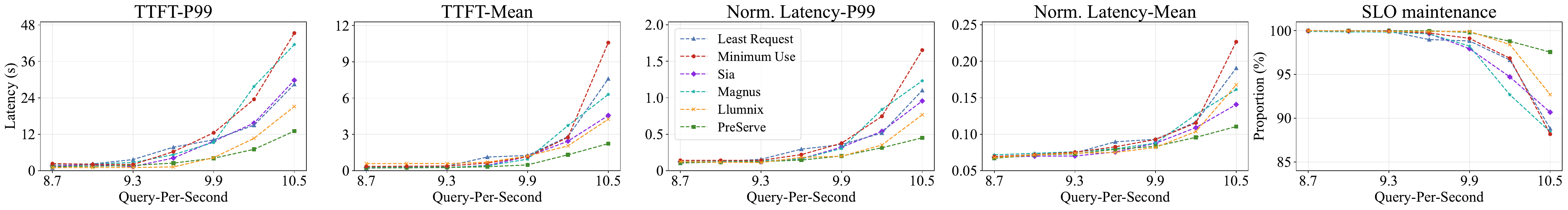}
    \caption{The latency and SLO maintenance of serving four LLM instances with different QPS under various routing strategies.}
    \label{fig:RQ3_request_routing}
\end{figure*}

\noindent
\add{\textbf{Settings.}}
In this research question (RQ), we evaluate the end-to-end performance of \nm in routing requests to balance instance loads and reduce serving latency.
Following previous studies~\cite{shen2021defuse,sun2024llumnix}, we utilize prompt-response pairs from the ShareGPT dataset~\cite{ShareGPT} to generate requests.
These requests are produced at varying query-per-second (QPS) rates, following a Poisson distribution over a fixed duration (\eg 10 minutes).

\noindent
\textbf{Baselines.}
\delete{We select three widely used load-balancing algorithms from traditional service systems as baseline methods~\cite{rasmussen2008round,zhu2018improved,singh2018improved}: \emph{round-robin (RR)}, \emph{least request (LR)}, and \emph{minimum use (MU)}. Detailed settings for these baselines are provided in Sec.~\ref{sec:motivation_opportunity_2}.}
\add{We select five widely-used and SOTA request routing strategies: (1) \emph{least request (LR)} and \emph{minimum use (MU)}: Two conventional load-balancing heuristics, and the detailed settings are provided in Sec.~\ref{sec:motivation_opportunity_2}. (2) \emph{Sia}~\cite{jayaram2023sia}: A scheduler for ML clusters that optimizes job placement and resource allocation. We adapt it to learn a goodput model (mapping request features to instance throughput) from historical data and then use an integer linear programming (ILP) solver to determine the optimal routing policy, based on its implementation~\cite{Sia-artifact}. (3) \emph{Magnus}~\cite{cheng2024enabling}: A LMaaS scheduling framework that leverages predicted request output lengths. It groups requests with similar predicted lengths to minimize padding overhead. (4) \emph{Llumnix}~\cite{LLumnix}: Following the RQ2 baseline setting, we evaluate LLumnix's request scheduling capability separately.}

\noindent
\add{\textbf{Results.}}
The experimental results are presented in Fig.~\ref{fig:RQ3_request_routing}.
We have made the following observations:
\begin{itemize}[leftmargin=*, topsep=0pt, parsep=0pt]
    \item When the QPS is low (\ie below 9.3), the system operates under a light load, and all routing strategies maintain comparable latency. 
    \item As QPS increases, simple strategies like LR, MU, and Magnus fail to balance the dynamic LLM workloads, leading to high tail latency. For instance, at 10.5 QPS, Magnus reaches a P99 normalized latency of 1.23s, violating SLOs for 11.56\% of requests. This is mainly because these methods are not designed for dynamic loads, and Magnus's routing logic is incompatible with modern continuous batching engines~\cite{yu2022orca}, where requests can dynamically join and leave batches.
    \item Sia's optimization-based routing policy, which lacks foresight into future loads, results in a P99 latency of 0.956s and only 90.68\% SLO adherence, as the QPS reaches 10.5. 
    \item Llumnix is the strongest baseline due to its ability to dynamically reschedule requests from overloaded to underloaded instances. However, this incurs substantial overhead from KV cache migration, leading to a considerable mean TTFT (4.26s), elevated normalized latency (0.767s), and only 92.7\% SLO adherence..
    \item In contrast, \nm utilizes a predictive, load-aware routing strategy that considers both current system status and future instance load. This approach allows \nm to maintain a low mean TTFT (2.24s), low normalized latency (0.45s), and a minimal SLO violation rate (2.44\%). These results correspond to performance improvements of 47.4\%, 41.3\%, and 66.58\%, respectively, over the best-performing baseline, Llumnix.
\end{itemize}

\subsection{RQ4: LMaaS Management Efficiency}

\noindent
\add{\textbf{Settings.}}
In this section, we evaluate the efficiency of \nm in managing LMaaS services, focusing on its ability to maintain low overhead and high scalability.
The primary cost associated with \nm arises from the load prediction for each request and the maintenance of load anticipators for each LLM instance.
To quantify this overhead, we measured the routing time for each LLM request under non-overloaded benchmark conditions, as specified in RQ3.
The management overhead is defined as the interval between the arrival of a request at the router and its assignment to a specific LLM instance, encompassing both response length prediction and the subsequent routing decision.
Moreover, we compared this overhead against metrics, including the average TTFT, normalized latency, and end-to-end latency across all served requests.

\begin{figure}[t]
    \centering
    \includegraphics[width=\columnwidth]{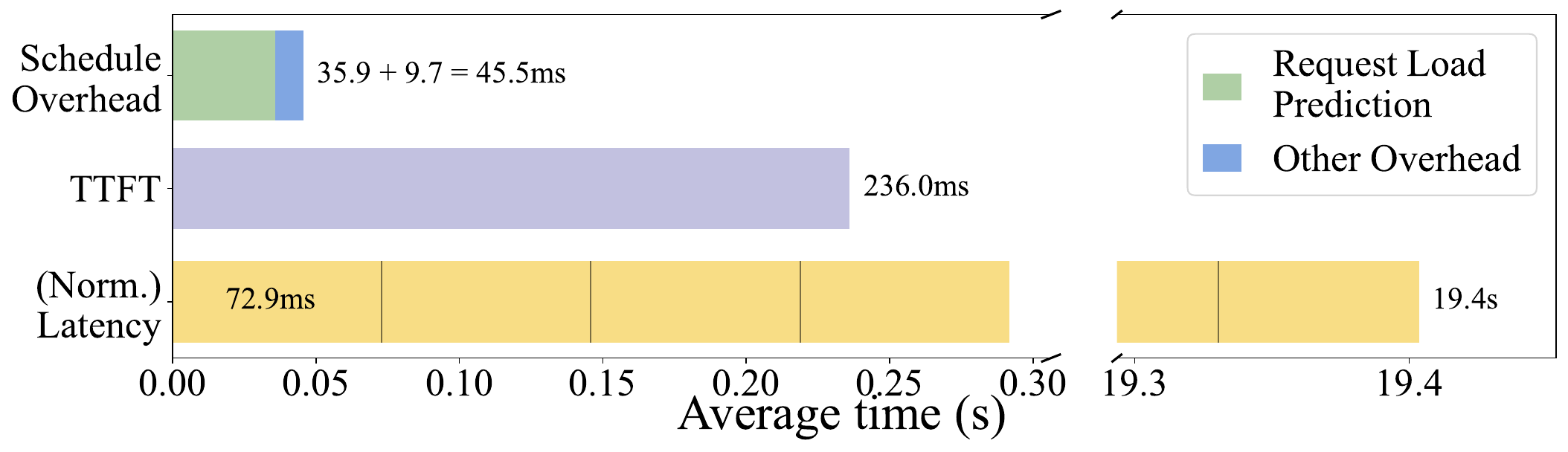}
    \caption{LMaaS management overhead of \nm.}
    \label{fig:RQ4_efficiency}
\end{figure}

\noindent
\add{\textbf{Results.}}
Fig.~\ref{fig:RQ4_efficiency} presents the aggregated average latency results for processed requests.
Our analysis reveals that load prediction is the primary part of management overhead, averaging 35.9 ms per request.
Other parts, such as load anticipator maintenance, collectively account for an additional 9.7 ms.
In comparison, the average first token latency (TTFT) and per-token latency (normalized latency) stand at 236 ms and 72.9 ms, approximately 5.2$\times$ and 1.6$\times$ greater than the overhead, respectively.
As depicted in Fig.~\ref{fig:RQ4_efficiency}, these latencies increase sharply to several thousand milliseconds and several hundred milliseconds when load imbalance occurs among instances.
Furthermore, the average end-to-end latency across all processed requests is 19.4 seconds, exceeding the scheduling overhead by three orders of magnitude.
Notably, \nm introduces an average additional latency of merely 0.23\% relative to the e2e request latency, yet it effectively mitigates latency spikes caused by load imbalance, underscoring its practical effectiveness.

\section{Threats to Validity}

\noindent
\textbf{Evaluated models and serving frameworks.}
We implemented and evaluated \nm with the LLaMA model family~\cite{LlamaFamily} and the vLLM framework~\cite{vllm}, given their popularity and extensive use in recent studies~\cite{sun2024llumnix,lazuka2024llm,hu2024memserve}.
However, \nm is designed to be non-intrusive and can be readily integrated with various LLMs and inference frameworks.
Furthermore, since \nm is not tied to specific hardware architectures and operates mainly at the software service level, its broad applicability is assured.
The core principles \nm relies on, such as autoregressive generation and iteration-based inference, are fundamental across diverse LLM frameworks, ensuring its effective generalization to various LMaaS platforms.

\noindent
\textbf{Reliance on Historical Request Data.}
\nm relies on historical LMaaS request data to train the hierarchical predictor and profile the serving capabilities of LLM instances.
The absence of such data at the initial stage of service deployment presents a ``cold start'' challenge. This limitation is, however, transient; the predictors require a minimal volume of training data, often corresponding to a few days of service operation. Throughout this preliminary phase, default scaling and routing protocols can be implemented pending the accumulation of adequate data. Furthermore, to accommodate the dynamic characteristics of the LLM service, the predictors can be periodically retrained.

\section{Related Work}

We structure our literature review in the following two domains:

\noindent
\textbf{(1) Service-level Management.}
Service management and SLO maintenance have been extensively studied in traditional cloud and web systems~\cite{guo2024pass,gambi2012modeling,chen2016self,qiu2020firm}, primarily focusing on two strategies: automated resource provisioning (\ie scaling)~\cite{vayghan2021kubernetes,karol2024self,song2024funcscaler,ahmad2025towards} and dynamic request routing (\ie load balancing)~\cite{bhattacharya2024dynamically}.

\noindent
\emph{(1.a) Instance Auto-scaling.} In traditional cloud services, predictive auto-scaling is a well-established technique, using workload forecasts to proactively provision resources~\cite{guo2024pass}. However, the significant instance startup times in LMaaS~\cite{fu2024serverlessllm} render purely reactive scaling ineffective. To address this, some recent works focus on vertical scaling; for example, µ-Serve~\cite{qiu2024power} dynamically adjusts GPU frequencies to handle minor load fluctuations. Other works employ reactive horizontal scaling. Llumnix~\cite{sun2024llumnix}, for instance, mitigates overload by migrating in-flight requests to new instances, though this incurs communication overhead. In contrast, \nm advances the state-of-the-art by introducing a hierarchical prediction mechanism. This allows for fine-grained load projections, enabling proactive horizontal scaling that preempts congestion and avoids the overhead of reactive migration.

\noindent
\emph{(1.b) Request Routing.}
Effective request routing is critical for performance. Early systems like DeepSpeed-MII~\cite{Deepspeed-mii} used simple round-robin policies. PiA~\cite{zheng2024response} and Magnus~\cite{cheng2024enabling}, for example, improve performance by grouping requests with similar predicted output lengths. However, this approach is often incompatible with the widely-used continuous batching technique~\cite{yu2022orca}. Other works like Sia~\cite{jayaram2023sia} apply optimization techniques, using ILP solvers to find optimal job placements in general ML clusters. \nm contributes a novel routing strategy that leverages its predictive hierarchy to intelligently dispatch requests, achieving superior load distribution while remaining compatible with modern serving frameworks.

\noindent
\textbf{(2) Instance-Level LLM Inference Optimization}
A distinct but complementary line of research focuses on optimizing the performance of a single LLM inference instance on fixed hardware. These works aim to boost throughput or reduce latency by optimizing the inference engine. For example, some methods disaggregate prefill and decode phases onto separate hardware (DistServe~\cite{zhong2024distserve}), enable dynamic sequence parallelism (LoongServe~\cite{wu2024loongserve}), or use hybrid CPU-GPU execution based on neuron activation locality (PowerInfer~\cite{song2024powerinfer}).
These works are orthogonal to \nm, as it operates at a higher service management layer, orchestrating entire instances, whereas these methods optimize operations \textit{within} an instance. As such, these approaches are complementary: the performance gains from an optimized inference engine can be further amplified by \nm's intelligent service-level scaling and routing.


\section{Conclusion}

In this paper, we introduce \nm, an LMaaS management framework that addresses key challenges in LMaaS management.
\nm employs hierarchical predictions, using a service workload predictor for periodic token density forecast and a request load predictor for per-request load estimation. 
By combining long- and short-term predictions, \nm proactively adjusts resource provision and optimizes request routing to balance instance loads. 
Evaluations on real-world LMaaS datasets show that \nm outperforms all baselines, significantly reducing latency and resource consumption with minimal overhead.
We believe \nm will contribute to advancing effective LLM service management.

\begin{acks}
This work was supported by the Research Grants Council of the Hong Kong Special Administrative Region, China (No. SRFS2425-4S03 of the Senior Research Fellow Scheme and No. CUHK 14209124 of the General Research Fund).
\end{acks}

\balance
\bibliographystyle{ACM-Reference-Format}
\bibliography{sample-base}

\end{document}